\documentclass[sigconf,10pt]{acmart}
\usepackage[nolist,nohyperlinks]{acronym}
\usepackage{listing}
\usepackage{listings}
\usepackage{listings-rust}
\usepackage{fancyvrb}
\usepackage{subfig}
\usepackage{semantic}
\newacro{FSM}{Finite State Machine}
\newacro{BNF}{Backus-Naur Form}
\newacro{AST}{Abstract Syntax Tree}
\newacro{DFS}{Depth First Traversal}
\newacro{CEC}{Columbia Esterel Compiler}
\newacro{CFG}{Control Flow Graph}
\newacro{GALS}{Globally Asynchronous Locally Synchronous}
\newacro{LTL}{Linear Temporal Logic}
\newacro{SOS}{Structural Operational Semantics}
\newacro{FSMD}{Finite State Machine with Data}
\allowdisplaybreaks

\AtBeginDocument{%
  }

\setcopyright{acmlicensed}
\copyrightyear{2018}
\acmYear{2018}
\acmDOI{XXXXXXX.XXXXXXX}

\acmConference[]{}{2025}{}
\acmISBN{978-1-4503-XXXX-X/18/06}

\begin{document}

\title{Efficient compilation and execution of synchronous programs
 via type-state programming}

\author{Avinash Malik}
\email{avinash.malik@auckland.ac.nz}
\orcid{0000--0002--7524--8292}
\affiliation{%
  \institution{Department of Electrical, Computer, and Software
Engineering, University of Auckland}
  \city{Auckland}
  \country{New Zealand}
}


\begin{abstract}
  Synchronous programs are used extensively in implementation of safety
  critical embedded software. Imperative synchronous programming
  languages model multiple \acfp{FSM} executing in lockstep at logical
  clock ticks. The synchronous view of time along with the \ac{FSM}
  based design enables easier formal verification. The synchronous
  composition of multiple \acp{FSM}, during compilation, results in the
  well known state space explosion problem. Hence, efficiently compiling
  imperative synchronous programs into small and fast executables is
  challenging. This paper introduces a novel linear time compilation
  technique for automata based compilation of synchronous programs.
  Graph based rewrite rules for kernel programming constructs are
  introduced. A linear time algorithm applies these rules to produce a
  \ac{FSM}. The \ac{FSM} is then encoded into a type-state program using
  template meta-programming in C++. Experimental results show that the
  compilation time and generated binary size is comparable, while the
  execution times are on average 31---60\% faster than current
  state-of-the-art compilers.
\end{abstract}





\maketitle

\section{Introduction}
\label{sec:introduction}

Synchronous programming is a popular
paradigm~\cite{halbwachs1992synchronous, benveniste2003synchronous} for
programming safety critical embedded
systems~\cite{bochot2009model,ferreira2023model,colacco2017scade,berry2000esterel}.
The primary motivation for development of synchronous programming
languages such as Esterel~\cite{berry1985esterel,boussinot1991esterel},
Lustre~\cite{halbwachs1991synchronous}, etc is that they provide a
logical model of time, which allows for formal reasoning about program
correctness. The Esterel~\cite{berry1999esterel} programming language in
particular is well suited for embedded programming as it models
\acfp{FSM} and it allows for easy interaction with procedural languages
such as C, being compiled into a C back-end. Individual Esterel
\texttt{modules} are \acp{FSM}. These \texttt{modules} can be executed
in synchronous parallel, akin to threads executing on a single (logical)
clock edge (tick). All parallelism in the described Esterel program is
compiled away to produce single threaded C code for execution.

Efficient compilation and execution of Esterel into C is
challenging~\cite{potop2007compiling}. Compiling multiple
\texttt{modules} executing in synchronous parallel requires synchronous
composition of multiple \acp{FSM} into a single \ac{FSM} for execution,
which leads to the well known state space explosion problem.
Constructing a single \ac{FSM} requires performing a cartesian product
of all states, in all modules of the program, which has an algorithmic
complexity of $O(\Pi_{i=1}^{M}N_i)$, where $M$ is the number of modules,
executing in synchronous parallel, in the program, and $N_{i}$ is the
number of states in each module. Hence, constructing a classical
synchronous composition in general takes an exponentially long time in
the number of state of the program.

Different approaches have been proposed for compiling imperative
synchronous programming languages like
Esterel~\cite{potop2007compiling,yip2023synchronous}. Three major
approaches to compilation are: \textcircled{1} generating a single
\ac{FSM}, after synchronous composition, in the form of a switch-case
statement in C --- called automata
compilation~\cite{berry1999constructive}. \textcircled{2} Generating a
control flow graph~\cite{edwards1999compiling} akin to traditional
compilation of programming languages, and \textcircled{3} Circuit based
compilation~\cite{berry1999esterel}, which first translates Esterel
programs into circuit equations, which get executed as C programs. All
compilation techniques have trade-offs. The executable code generated
from the automata compiler is very fast. However, the generated code
size is too large and the compilation might not finish, because of the
state space explosion problem~\cite{edwards2007code}. The control flow
graph based compilation techniques produce small sized binaries, with
fast compilation time and efficient execution. However, program
analysis, for checking program properties, becomes difficult. Circuit
based approach is supported by the official Esterel compiler. However,
it has been shown to produce slow runtimes~\cite{edwards2007code}.

It is an open challenge whether a compilation technique exists that can
quickly compile imperative synchronous programs into efficient, in size
and runtime, executables.

This compilation challenge is addressed in this work. The main
\textbf{contributions} of the paper are: \textcircled{1} new syntactic
graph rewrite rules for Esterel. 
\textcircled{2} These translation rules enable compositional and linear
time compilation of Esterel program into a \acf{FSM}. \textcircled{3}
The generated \ac{FSM} is then encoded into a
type-state~\cite{garcia2014foundations,strom1986typestate} program in
C++. To the best of our knowledge we are the first to present a linear
time \ac{FSM} compilation approach for imperative synchronous programs.
Experimental results show that: \textcircled{a}, the compile times
comparable to the current state-of-the-art Esterel compilers.
\textcircled{b} The generated binary size is comparable to the current
state-of-the-art. Finally, \textcircled{c} the runtime is 31--60\%
\textit{faster} than the current state-of-the-art.



The rest of the paper is arranged as follows:
Section~\ref{sec:intr-type-state} introduces type-state programming,
followed by the \textbf{major technical contributions} in
Sections~\ref{sec:kern-synchr-lang} and~\ref{sec:comp-kern-synchr}. A
kernel imperative synchronous programming language with its syntax and
semantics is presented in Section~\ref{sec:kern-synchr-lang}, followed
in Section~\ref{sec:comp-kern-synchr}, with a systematic syntax-driven
translation technique for compiling the kernel language into a \ac{FSM},
accompanied with a linear time compilation algorithm. Experimental
results comparing the presented compilation technique with current
state-of-the-art compilers for Esterel program are presented in
Section~\ref{sec:experimental-results}. Related work is presented in
Section~\ref{sec:related-work}. The paper concludes in
Section~\ref{sec:concl-future-work}.

\section{Introduction to type-state programming in C++}
\label{sec:intr-type-state}

\begin{lstlisting}[language=C++, numbers=left,
  escapechar=|,basicstyle=\scriptsize,
  caption={A type-state Traffic light controller in C++}, label={lst:1}]
  #include <variant>
  extern "C" unsigned char timerEvent();
  extern "C" void setOutput(unsigned char);
  // XXX: The generic state|\label{ln:1}|
  struct State {};|\label{ln:2}|
  // XXX: The different states the traffic light can be in
  struct RED : State {};|\label{ln:3}|
  struct GREEN : State {};|\label{ln:4}|
  struct YELLOW : State {};|\label{ln:5}|
  // XXX: The controller object templated over generic State
  template <typename State> struct Con {};|\label{ln:6}|
  // XXX: The possible states the controller can be in
  using ConState = std::variant<Con<RED>, Con<GREEN>,
  Con<YELLOW>>; |\label{ln:7}|
  // XXX: The FSM transition functions
  // specialised for each state
  constexpr ConState Con<RED>::tick() { |\label{ln:8}|
    // XXX: Once some external timer event happens, go to
    // next state. 
    // set output from controller when going to next state
    return timerEvent() ? setOutput(1), Con<GREEN>{} :
    Con<RED>{};
  }|\label{ln:9}|
  constexpr ConState Con<GREEN>::tick() {|\label{ln:10}|
    return timerEvent() ? setOutput(2), Con<YELLOW>{} :
    Con<GREEN>{};
  }|\label{ln:11}|
  constexpr ConState Con<YELLOW>::tick() {|\label{ln:12}|
    return timerEvent() ? setOutput(0), Con<RED>{} :
    Con<YELLOW>{};
  }|\label{ln:13}|
  int main(void) {|\label{ln:14}|
    // XXX: Initial state of the controller
    ConState st = Con<RED>{}; |\label{ln:15}|
    setOutput(0); //Set controller output to 0 (state RED)
    // XXX: Ticks from RED --> GREEN --> YELLOW -->....
    st = std::visit([](auto &&t) {return t.tick();},
    std::move(st)); |\label{ln:16}|
  }
\end{lstlisting}

Type-state programming~\cite{garcia2014foundations,strom1986typestate}
is the paradigm of encoding the state of an imperative program within
the type system of a programming language. Type-state programming
enables localised reasoning and compositional verification of imperative
programs. We want to leverage the compositional nature of type-state
programming for linear time compilation of imperative Esterel programs
into \acp{FSM}.

Listing~\ref{lst:1} shows a pedagogical example of an embedded
controller, controlling a traffic light, designed using the type-state
programming paradigm in C++. The program declares the three states
(lines~\ref{ln:3}---\ref{ln:5}) of the traffic light. The controller
\texttt{Con} (line~\ref{ln:6}) is templated over the generic
\texttt{State} type. There are three transition functions for the
controller (lines~\ref{ln:8}---\ref{ln:13}). These transition functions
are specialised for each state template parameter. Hence, if the
controller is currently in state \texttt{RED}, then the controller can
remain in that state or go to state \texttt{GREEN} after some timeout
happens (lines~\ref{ln:8}-\ref{ln:9}), no other state is reachable from
\texttt{RED}. Similar functionality is encoded for states \texttt{GREEN}
(lines~\ref{ln:10}-\ref{ln:11}) and \texttt{YELLOW}
(lines~\ref{ln:12}-\ref{ln:13}). The \texttt{main} function initialises
the controller to the \texttt{RED} state. Then an infinite number of
ticks happen with the state of the controller changing according to the
transitions functions (line~\ref{ln:16}).

The main points of note are: \textcircled{1} the states of the \ac{FSM}
are encoded into types using \texttt{struct}. \textcircled{2} The
controller's functionality is divided into different functions
\textit{dependent} upon the type (state) the controller is in. For
example, on lines~\ref{ln:8}-\ref{ln:9}, when the controller is in state
\texttt{RED}, we can reason \textit{locally} that the controller can
only remain in this state or go to state \texttt{GREEN}. This allows for
modular formal reasoning and contract based verification using the type
system~\cite{field2003typestate,garcia2014foundations}. \textcircled{3}
States are empty \texttt{struct}(s) (lines~\ref{ln:3}-\ref{ln:5}).
Hence, there is no need to store the state in a vector, reducing memory
footprint. \textcircled{4} Finally, the \texttt{constexpr} functionality
of C++ allows for compile time execution and aggressive optimisations,
which is leveraged for efficient executable code generation. 

We will leverage type-state programming principles to produce efficient
executable binaries with fast execution times.

\section{Kernel synchronous language --- syntax and semantics}
\label{sec:kern-synchr-lang}

We first describe the syntax and semantics of a kernel synchronous
programming language. For sake of clarity, we only show prominent kernel
statements. The developed compiler supports a number of other constructs
such as data transformations, derived constructs, type inferred
functions, and integration with C in addition to these kernel
statements.

\subsection{Syntax of the kernel language}
\label{sec:syntax}

The \ac{BNF} grammar of the kernel language is shown in
Figure~\ref{fig:1}. There are 8 kernel statements. The first statement
declares a signal, which can be optionally an \texttt{input} or
\texttt{output} signal interacting with the environment. The
\texttt{nothing} statement performs no operation. The \texttt{pause}
construct waits until a logical tick (clock edge) occurs. \texttt{Pause}
is the only statement that can introduce a state in the program.
Construct \texttt{emit} produces an output to the environment. Two
statements can be executed sequentially using the semicolon
(\texttt{;}). Loops are always infinite and there needs to exist at
least a single \texttt{pause} in all paths of the loop body. A program
can branch using the \texttt{if-else} construct checking a logical
signal expression. The \texttt{abort} construct pre-empts the execution
of its body if the logical signal expression is true. Finally, two or
more blocks can be executed in parallel synchronously (executing at the
same clock tick, in lockstep) using the parallel construct
(\texttt{||}).

\begin{figure}[tbh]
  \centering
\begin{verbatim}
stmt := [input | output] signal <name>
        |nothing | pause | emit <signal>
        |stmt; stmt | loop{stmt}
        |if(<expr>) {stmt} else {stmt}
        |abort(<expr>) {stmt}
        |{stmt}||{stmt}||{stmt}||...
\end{verbatim}
  \caption{The \ac{BNF} grammar of kernel synchronous language}
  \label{fig:1}
\end{figure}

\subsection{Informal semantics of the kernel language}
\label{sec:semantics}

Figure~\ref{fig:2} shows the \texttt{ABRO} running example used to
describe the semantics and compilation strategy in the rest of the
paper. Lines~\ref{re:ln:1} and~\ref{re:ln:2} define three input signals
and one output signal used to interact with the environment. The program
runs infinitely (loop on lines~\ref{re:ln:3}---\ref{re:ln:11}), reacting
to inputs from the environment. The program waits for signals \texttt{A}
and \texttt{B} to be present (lines~\ref{re:ln:5} and~\ref{re:ln:7}) in
parallel (line~\ref{re:ln:6}). Labels \texttt{S0} and \texttt{S1} on
lines~\ref{re:ln:5} and~\ref{re:ln:7} show the \ac{FSM} states waiting
for the input signals. Once these signals are present, the synchronous
parallel statement (line~\ref{re:ln:6}) finishes and emits signal
\texttt{O} to the environment (line~\ref{re:ln:8}). The program then
halts on \ac{FSM} state \texttt{S2}, waiting for signal \texttt{R} to be
present (line~\ref{re:ln:9}). When signal \texttt{R} is present, at any
state in the program, the program is reset
(line~\ref{re:ln:4}---\ref{re:ln:10}).

\begin{figure}[tb]
  \centering
\begin{Verbatim}[numbers=left,numbersep=3pt,numberblanklines=false,fontsize=\small,
  commandchars=\\\[\]]
input signal A, B, R; //Input signals from environment \label[re:ln:1]
output signal O; //Output signal to environment\label[re:ln:2]
loop { \label[re:ln:3]
 abort(R) { //abort body when signal R is present \label[re:ln:4]
  {abort(A){loop{S0: pause}}} //stmt-1--wait  signal A \label[re:ln:5]
  || //run stmt-1 and stmt-2 in synchronous parallel \label[re:ln:6]
  {abort(B){loop{S1: pause}}}; //stmt-2--wait  signal B \label[re:ln:7]
  emit O; // emit O if A and B and not R \label[re:ln:8]
  loop{S2: pause} //halt \label[re:ln:9]
 } \label[re:ln:10]
} //restart (loop-back) program when R is present \label[re:ln:11]
\end{Verbatim}
  \caption{The \texttt{ABRO} running example}
  \label{fig:2}
\end{figure}

\begin{figure}[tbh]
  \centering
  \includegraphics[scale=0.8]{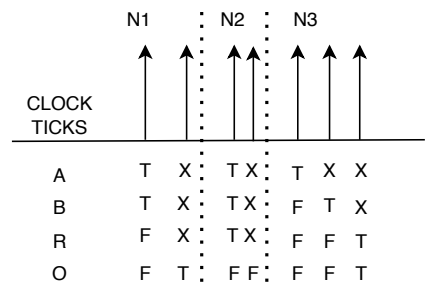}
  \caption{Timing behaviour for the \texttt{ABRO} running example.
    \texttt{T} and \texttt{F} indicate True and False, \texttt{X}
    indicates don't care condition.}
  \label{fig:3}
\end{figure}

Figure~\ref{fig:3} shows a timing diagram with different scenarios for
the \texttt{ABRO} running example in Figure~\ref{fig:2}. In scenario N1,
signals \texttt{A} and \texttt{B} are present together in the first
clock tick, with signal \texttt{R} being absent. Hence, signal
\texttt{O} is emitted in the second (next clock tick). In scenario N2,
signals \texttt{A}, \texttt{B}, and \texttt{R} are all present at the
same time in the first tick. Hence, signal \texttt{O} is \textit{not}
emitted in the next tick, because the program is reset due to presence
of signal \texttt{R}. Finally, in scenario N3, signals \texttt{A} and
\texttt{B} are present in the first and second tick, respectively. In
the third tick, signal \texttt{O} is emitted, even with the presence of
signal \texttt{R}, because the program reacts to the (Boolean) status of
signals from the previous tick only.

The compilation strategy will construct a \ac{FSM} representation of the
synchronous program in linear time. A type-state program will be
generated at the back-end from the constructed \ac{FSM}. We first
describe the formal semantics of the kernel language, followed by the
compilation strategy derived from these formal semantics.

\subsection{Formal semantics of the kernel statements}
\label{sec:form-semant-kern}

We give a \acf{SOS} of the kernel language. For every syntactic kernel
construct, we have a rewrite rule. The state of the program consist of
signal statuses. The signal statuses are captured in the maps $I$ and
$O$, for input, output signals, respectively. The status of the internal
signals are kept in map $S$. The statuses of the signals in the previous
tick are also kept within their individual signal maps.

The rewrite rules are presented as inference rules from
Equations~(\ref{eq:nothing})--(\ref{eq:par4}). An inference rule consists
of an antecedent and a consequent represented as
$\frac{antecedent}{consequent}$. If the antecedent holds, then the
consequent follows. An empty antecedent indicates \texttt{True}
($\top$). The antecedent and consequent rules are written using the
relational operator
$I,O,S,construct \\ \xrightarrow{tick} I',O',S'construct'$, where the
left side of the arrow is the syntactic construct and the right side of
the arrow is the resultant rewrite. The rewrite rules (may) also change
the signal statuses of the program. We write $O[K]$, for the output map
$O$ and for some signal $K$ in that map to lookup the status of the
signal. Similarly, $O[K] \leftarrow \top$ indicates setting the status of the signal
in the map to true. Same for input and local signal maps $I$ and $S$.
Finally, $I, O, S |- \mathrm{expr}$ indicates checking the status of a
signal expression within the context of the previous tick' signal status
in the maps.

The $tick$, atop the rewrite rule, indicates logical tick completion and
generation of \ac{FSM} state (c.f. Section~\ref{sec:syntax}). The
variable $tick \in \{0, 1\}$, takes a value of 1, if a program (or its
constituent) completes a logical tick and generates a \ac{FSM} state,
else it takes a value of 0.

\vspace{-8mm}

\begin{small}
  \begin{gather}
    \inference{}{I,O,S,\mathtt{nothing} \xrightarrow{0} I,O,S,\mathtt{nothing}}
    \label{eq:nothing}
    \\
    \inference{K\in O}
    {I,O,S,\mathtt{emit\ K} \xrightarrow{0} I, O[K]\leftarrow \top,S,\mathtt{nothing}}
    \label{eq:emit}
    \\
    \inference{K\in S}
    {I,O,S,\mathtt{emit\ K} \xrightarrow{0} I, O, S[K]\leftarrow \top,\mathtt{nothing}}
    \label{eq:emiti}
    \\
    \inference{}{I,O,S,\mathtt{input\ signal\ K} \xrightarrow{0}
      I[K]\leftarrow \mathtt{inst}(I,K),O,S,\mathtt{nothing}}
    \label{eq:inputsig}
    \\
    \inference{}{I,O,S,\mathtt{output\ signal\ K} \xrightarrow{0}
      I,O[K]\leftarrow \mathtt{inst}(O,K),S,\mathtt{nothing}}
    \label{eq:outsig}
    \\
    \inference{}
    {I,O,S,\mathtt{signal\ K} \xrightarrow{0} I, O, S[K]\leftarrow \mathtt{inst}(S,K),
      \mathtt{nothing}}
    \label{eq:sig}
    \\
    \inference{}{I,O,S,\mathtt{pause} \xrightarrow{1} I,O,S,\mathtt{nothing}}
    \label{eq:pause}
    \\
    \inference{I,O,S,\mathtt{s1} \xrightarrow{1} I',O',S',\mathtt{s1'}}
    {I,O,S,\mathtt{s1;s2} \xrightarrow{1} I',O',S',\mathtt{s1';s2}}
    \label{eq:seq1}
    \\
    \inference{I,O,S,\mathtt{s1'} \xrightarrow{0} I',O',S',\mathtt{nothing}}
    {I,O,S,\mathtt{s1';s2} \xrightarrow{0} I',O',S',\mathtt{s2}}
    \label{eq:seq2}
    \\
    \inference{I,O,S |- \mathtt{expr} => \top}
    {I,O,S,\mathtt{if(expr)\ {p}\ else\ {q}} \xrightarrow{0} I,O,S,\mathtt{p}}
    \label{eq:ift}
    \\
    \inference{I,O,S |- \mathtt{expr} => \bot}
    {I,O,S,\mathtt{if(expr)\ {p}\ else\ {q}} \xrightarrow{0} I,O,S,\mathtt{q}}
    \label{eq:ife}
    \\
    \inference{}
    {I,O,S,\mathtt{loop\{p\}} \xrightarrow{0} I,O,S,\mathtt{p;loop\{p\}}}
    \label{eq:loop}
    \\
    \inference{I, O, S |- \mathtt{expr} => \top}
    {I,O,S,\mathtt{abort(expr)\{p\}} \xrightarrow{0} I',O',S',\mathtt{nothing}}
    \label{eq:abort1}
    \\
    \inference{I, O, S |- \mathtt{expr} => \bot \\
      I, O,S, \mathtt{p} \xrightarrow{1} I', O',S', \mathtt{p'}}
    {I,O,S,\mathtt{abort(expr)\{p\}} \xrightarrow{1}
      I',O',S',\mathtt{abort(expr)\{p'\}}}
    \label{eq:abort2}
    \\
    \inference{I, O, S |- \mathtt{expr} => \bot \\
      I, O,S, \mathtt{p} \xrightarrow{0} I', O',S', \mathtt{nothing}}
    {I,O,S,\mathtt{abort(expr)\{p\}} \xrightarrow{0} I',O',S',\mathtt{nothing}}
    \label{eq:abort3}
    \\
    \inference{I, O,S, \mathtt{p} \xrightarrow{0} I',O',S',\mathtt{nothing} \\
      I,O,S,\mathtt{q} \xrightarrow{0} I'',O'',S'',\mathtt{nothing}}
    {I,O,S,\mathtt{p||q} \xrightarrow{0} I'\cup I'',O'\cup O'',S'\cup S'',\mathtt{nothing}}
    \label{eq:par1}
    \\
    \inference{I, O,S, \mathtt{p} \xrightarrow{1} I'',O'',S'',\mathtt{p'} \\
      I,O,S,\mathtt{q} \xrightarrow{0} I',O',S',\mathtt{nothing}}
    {I,O,S,\mathtt{p||q} \xrightarrow{1} I'\cup I'',O'\cup O'',S'\cup S'',\mathtt{p'||nothing}}
    \label{eq:par2}
    \\
    \inference{I, O,S, \mathtt{q} \xrightarrow{1} I'',O'',S'',\mathtt{q'} \\
      I,O,S,\mathtt{p} \xrightarrow{0} I',O',S',\mathtt{nothing}}
    {I,O,S,\mathtt{p||q} \xrightarrow{1} I'\cup I'',O'\cup O'',S'\cup S'',\mathtt{nothing||q'}}
    \label{eq:par3}
    \\
    \inference{I, O,S, \mathtt{p} \xrightarrow{1} I'',O'',S'',\mathtt{p'} \\
      I,O,S,\mathtt{q} \xrightarrow{1} I',O',S',\mathtt{q'}}
    {I,O,S,\mathtt{p||q} \xrightarrow{1} I'\cup I'', O'\cup O'',S'\cup S'',\mathtt{p'||q'}}
    \label{eq:par4}
  \end{gather}
\end{small}

The \texttt{nothing}, \texttt{emit} and signal declaration statements
execute instantaneously, without generating an internal \ac{FSM} state.
Hence, their rewrite rules (Equations~(\ref{eq:nothing})---(\ref{eq:sig}))
set the $tick$ value to 0. The signal declaration rules set the status
of the signal using the function $\mathtt{inst}(M, S)$ defined as:
\texttt{if M[S] then M[S] else false}; in their respective maps
(Equations~(\ref{eq:inputsig})---(\ref{eq:sig})). The $\mathtt{inst}$
function guarantees that if the signal is re-declared, possibly in a
\texttt{loop}, then the status of the signal is correctly updated. The
\texttt{emit} construct on the other hand sets the status of the signal
($K$) to true ($\top$); Equations~(\ref{eq:emit}) and~(\ref{eq:emiti})
depending upon the conditional that signal $K$ is an output signal or
local signal. All these instantaneous statements get rewritten into
\texttt{nothing} upon execution.

The \texttt{pause} construct is the only construct that introduces a
\ac{FSM} state in the program --- indicating completion of a logical
tick. Hence, its rewrite rule sets the value of the $tick$ to 1
(Equation~(\ref{eq:pause})). No change in signal statuses take place and
the construct is rewritten into \texttt{nothing} upon execution.

The sequential composition of kernel constructs follows two rewrite
rules (Equations~(\ref{eq:seq1}) and~(\ref{eq:seq2})). If the first
statement $\mathtt{s1}$ in the sequence $\mathtt{s1;s2}$ finishes with a
$tick$ value of 1, then the whole sequential composition \textit{pauses}
(Equation~(\ref{eq:seq1})). If the first statement finishes
instantaneously, then the sequential statement is rewritten into the
second construct $\mathtt{s2}$ (Equation~(\ref{eq:seq2})). Consider the
sequential composition of two statements: \texttt{pause;emit A}. The
program execution following the rewrite rules is shown in the derivation
tree in Equation~(\ref{eq:deriv}). First the \texttt{pause}
(Equation~(\ref{eq:pause})) and the first sequential rewrite rule
(Equation~(\ref{eq:seq1})) are applied resulting in $tick$ taking a
value of 1. Next, the \texttt{nothing} (Equation~(\ref{eq:nothing}))
rewrite rule along with the second sequential rewrite rule
(Equation~(\ref{eq:seq2})) results in sequential construct being
rewritten into just \texttt{emit A}. Finally, the \texttt{emit} rewrite
rule (Equation~(\ref{eq:emit})) is applied to get the signal \texttt{A}
status being set in the output map.

\vspace{-20pt}

\begin{gather}
  \inference{
    \inference{I, O, S, \mathtt{pause} \xrightarrow{1}
      I, O,S, \mathtt{nothing}}
    {I, O, S, \mathtt{pause;emit\ A} \xrightarrow{1}
      I,O,S,\mathtt{nothing;emit\ A}} \nonumber \\
    \vdots
    \nonumber \\
    \inference{I,O, S,\mathtt{nothing} \xrightarrow{0}
      I,O, S,\mathtt{nothing}}
    {I,O,S,\mathtt{nothing;emit\ A} \xrightarrow{0} I,O,S,\mathtt{emit\ A}}
  }{I,O,S,\mathtt{emit\ A} \xrightarrow{0} I,O[A]\leftarrow\top, S,\mathtt{nothing}} \\
  \label{eq:deriv}
\end{gather}

The two \texttt{if-else} rewrite rules
(Equations~(\ref{eq:ift})---(\ref{eq:ife})), correspond to the program
entering the \texttt{then} and \texttt{else} branches depending upon the
evaluation of the logical signal \texttt{expr}. The \texttt{loop}
rewrite (Equation~(\ref{eq:loop})) just unrolls the loop forever. There
are three rewrite rules for the \texttt{abort} construct. The first
rewrite (Equation~(\ref{eq:abort1})), states that if the \texttt{abort}
expression evaluates to true ($\top$), then the \texttt{abort} body is
immediately pre-empted and the statement is rewritten into
\texttt{nothing}. The second rewrite rule (Equation~(\ref{eq:abort2})),
states that; if the \texttt{abort} expression evaluates to false
($\bot$), and the \texttt{abort} body pauses producing a $tick$ value of
1, then the whole \texttt{abort} statement also pauses. Finally, the
third rewrite rule (Equation~(\ref{eq:abort3})), states that; if the
\texttt{abort} expression evaluates to false ($\bot$), and the abort
body finishes instantaneously, then the whole \texttt{abort} statement
also finishes instantaneously and gets rewritten into \texttt{nothing}.

The synchronous parallel construct ($\mathtt{p}||\mathtt{q}$) has four
rewrites (Equations~(\ref{eq:par1})---(\ref{eq:par4})). If both the
constituents of the parallel construct finish instantaneously, then the
whole parallel construct finishes instantaneously
(Equation~(\ref{eq:par1})). If the first thread pauses with a $tick$
value of 1, and the second finishes instantaneously, then the parallel
construct also pauses (Equation~(\ref{eq:par2})).
Equation~(\ref{eq:par3}) is the dual of the second rewrite. Finally, if
both the threads of the parallel construct pause then the whole parallel
construct also pauses (Equation~(\ref{eq:par4})). In all these
synchronous parallel rewrite rules, the updates to inputs and outputs
from both the threads are merged.

In addition to these rewrite rules, some book-keeping also takes place
when the program finishes its logical tick. Primarily, all the current
signal statuses are copied into their previous signal status in the
signal maps. The current signal statuses are set to false.

\section{Compilation of Kernel synchronous language to a \acs{FSM}}
\label{sec:comp-kern-synchr}

A systematic approach to compilation is necessary to reason about its
correctness. Hence, we take a syntax based rewrite approach to
compilation. First, for every syntactic construct (Figure~\ref{fig:1}),
an equivalent Graph rewrite is presented. These rewrites are derived
from the formal semantics defined in Section~\ref{sec:form-semant-kern}.
Next, an algorithm that traverses the \ac{AST} and applies the syntactic
rewrite rules is presented. These graph rewrite rules and the associated
compilation technique presented in this section are the major
contributions towards an efficient compilation technique for imperative
synchronous programs.

\subsection{Syntactic graph rewrite rules}
\label{sec:synt-transl-rules}

\begin{figure*}[tbh]
  \centering
  \subfloat[Rewrite for \texttt{emit}, \texttt{nothing}, and signal
declarations\label{fig:4a}]{\includegraphics[scale=0.6]{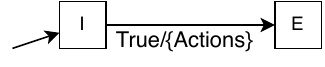}}
\hspace{12pt}
  \subfloat[Rewrite for
\texttt{pause}\label{fig:4e}]{\includegraphics[scale=0.6]{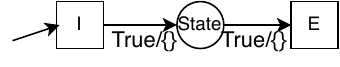}}
\hspace{12pt}
  \subfloat[Rewrite for sequential (\texttt{;}) and \texttt{loop}
  \label{fig:4b}]{\includegraphics[scale=0.6]{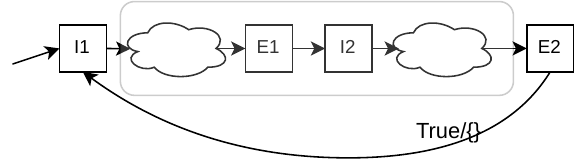}}

  \subfloat[Rewrite for presence
check\label{fig:4c}]{\includegraphics[scale=0.6]{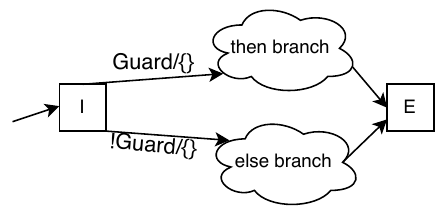}}
  \subfloat[Rewrite for
\texttt{abort}\label{fig:4d}]{\includegraphics[scale=0.65]{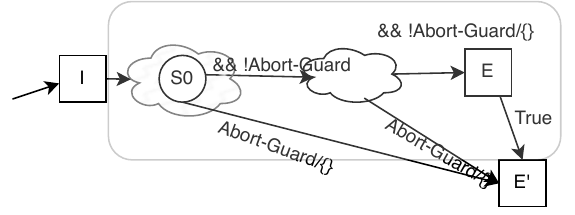}}
  \subfloat[Rewrite for synchronous parallel
(\texttt{||})\label{fig:4f}]{\includegraphics[scale=0.65]{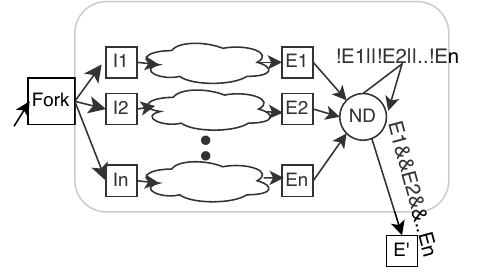}}

\caption{Syntactic graph rewrite rules for kernel statements}
  \label{fig:4}
\end{figure*}

Every syntactic construct is encoded as a graph. These rewrites are
shown in Figure~\ref{fig:4}. There are two types of nodes in the
generated graph: \textcircled{1} dummy nodes --- represented as squares
and \textcircled{2} state nodes --- represented as circles. Dummy nodes
indicate the start and end of individual instantaneous statements.
Program control-flow does \textit{not} stop (indicate end of current
logical tick) at these dummy nodes. These nodes help in compositional
construction of the program \ac{FSM}. The state nodes indicate end of a
clock tick and start of the next tick. The edges connecting the graph
are labelled with $\frac{guard}{action}$. The $guard$ represents the
signal expressions in branching and pre-emptive constructs such as
\texttt{if-else} and \texttt{abort}. Actions are the updates to signals
such as those performed by \texttt{emit}, etc.

Figure~\ref{fig:4a} gives the rewrite for the instantaneous
\texttt{emit}, \texttt{nothing}, and signal declaration statements.
These statements are always executed---antecedent of
Equations~(\ref{eq:nothing})---~(\ref{eq:sig}). Hence, the guard is just
\texttt{True} with their associated action sets as represented in the
consequent of the rewrite semantics
(Equations~(\ref{eq:nothing})---(\ref{eq:sig})).

The \texttt{pause} construct introduces a state node as shown in
Figure~\ref{fig:4e}. For example, the \texttt{S0:pause} construct on
line~\ref{re:ln:5} in the \texttt{ABRO} running example in
Figure~\ref{fig:2} would produce the graph in Figure~\ref{fig:4e}, with
circled (state) node labelled \texttt{S0}. While executing the generated
graph (\ac{FSM}) the program stops at these state nodes --- indicating the
end of the current logical clock tick. In the \textit{next} clock tick,
progress is made out of the state node (c.f. Equation~(\ref{eq:pause})
and e.g. Equation~(\ref{eq:deriv})).

Sequential construct (\texttt{;}) connects the last node \texttt{E1} to
the first nodes of the next construct \texttt{I2}
(Equations~(\ref{eq:seq1})---(\ref{eq:seq2})). The \texttt{loop}
construct builds an edge between the current last node in the graph to
the first node in the graph (Equation~(\ref{eq:loop})). The guards on
these edges are \texttt{True} (Figure~\ref{fig:4b}) with an empty action
set. The \texttt{if-else} construct builds a graph with two branches
(Figure~\ref{fig:4c}), with the logical signal expression guard and its
negation, respectively (c.f. Equations~(\ref{eq:ift})---(\ref{eq:ife})).
These branches are connected to the \texttt{then} and \texttt{else}
bodies, respectively.

Handling \texttt{abort} is more intricate (c.f. Figure~\ref{fig:4d}).
First initial (\texttt{I}) and end dummy nodes (\texttt{E'}) are
introduced. The initial dummy node is connected to the body of the
\texttt{abort} statement. Next, each \textit{state} node in the body of
\texttt{abort} is considered (e.g. \texttt{S0} in Figure~\ref{fig:4d}).
The outgoing edge guard from any state node, in the body, is anded with
the negation of the abort signal expression. This guard condition
modification ensures that the body of \texttt{abort} can proceed only if
the abort signal expression does \textit{not} hold
(Equations~(\ref{eq:abort2}) and~(\ref{eq:abort3})). Extra edges are
introduced from the state nodes to the end node \texttt{E'} guarded by
the abort signal expression. Hence, from any state, if the abort signal
expression holds, the \texttt{abort} body will be immediately exited
(Equation~(\ref{eq:abort1})).

The synchronous parallel construct (\texttt{||}) produces the most
complex graphs (c.f. Figure~\ref{fig:4f}). All $n$ statements executing
in synchronous parallel are handled individually. First the sub-graphs
for each of these individual parallel blocks are constructed. These are
shown in Figure~\ref{fig:4f} with their individual starting and end
nodes with \texttt{I1} through to \texttt{In} and \texttt{E1} through to
\texttt{En}, respectively. The synchronous parallel construct then
introduces three new nodes. Two dummy nodes \texttt{Fork} and
\texttt{E'} and a state node \texttt{ND}. The \texttt{Fork} node is
connected to the initial nodes of each of the parallel blocks. The state
node (\texttt{ND}) completes the execution of the parallel construct by
progressing to \texttt{E'} \textit{if and only if} all parallel
statements have reached their individual end nodes
(Equation~(\ref{eq:par1})). Else, the self-edge on \texttt{ND} is taken
(Equations~(\ref{eq:par2})---(\ref{eq:par4})). These graph rewrites are
combined together in an algorithm to generate the \ac{FSM}.


\subsection{Linear time algorithm to generate the program \ac{FSM}}
\label{sec:algor-gener-graph}

Listing~\ref{lst:2} gives the pseudo-code applying and combining the
graph rewrite to generate a \ac{FSM} from the \ac{AST} of a synchronous
program. We will use the \texttt{ABRO} running example to describe how
the algorithm works.

\begin{lstlisting}[language=Rust, numbers=left, escapechar=!,
  basicstyle=\scriptsize, caption={Rust compiler pseudo-code applying
    rewrite rules}, label={lst:2}]
  type Idx = (usize, usize);
  pub fn rewrite (stmt : &Stmt, narena: &mut Vec<GraphNode>,
  othreads: &mut Vec<Idx>) -> Idx {
    match stmt {
      Stmt::SignalDecl(io, s) !\label{ss}!
      | Stmt::Emit(s) | Stmt::Nothing =>
      {let i = buildNode(narena);
        let u = buildNode(narena); addGs(True);
        addAs(Some(s)); link(i,u);
        return (i, u);}
      Stmt::Pause(l) => {let i = buildNode(narena);
        let s = buildNode(l,narena);
        let u = buildNode(narena);
        addGs(s); addAs(s);
        link(s,i,u);
        return (i, u);} !\label{se}!
      Stmt::IfElse(s, tb, eb) => {
        let (it, et) = rewrite(tb, narena othreads);
        let (ie, ee) = rewrite(eb, narena, othreads);
        let i = buildNode();
        let e = buildNode(); addGs(s); link(i, it, ie);
        link(e, et, ee); return(i, e);}
      Stmt::Loop(body) =>  {
        let (i, e) = rewrite(body, narena, othreads); !\label{loops}!
        link(i, e); 
        return (i, e);} !\label{loope}!
      //rewrite s2 then s1
      Stmt::Seq(s1,s2) => { !\label{seqs}!
        let(i2,e2) = rewrite(s2, narena, othreads);
        let (i1, e2) = rewrite(s1, narena, othreads);
        link(i1, i2); link(e1, e2); return (i1, e2);} !\label{seqe}!
      Stmt::Abort(s, b) => {let (ib, eb) = rewrite(b, narena,
        othreads);
        addGuardNeg(narena, ib, eb, s);
        let e = buildNode(narena);
        addGuardEdges(narena, ib, eb, s, e);
        return(ib, e);}
      Stmt::Par([b]) => {
        let [(ic, ec)] = &b.map(|x| rewrite(x, narena,
        othreads));
        let (i, nd, e) = (buildNode(narena),
        buildNode(narena),
        buildNode(narena)); 
        buildGs(&ec, nd); buildGs(nd, e);
        link(i, &ic); link(&ec, nd); link(nd, e);
        othreads.push(ic); othreads.push(ec);
        return (i, e);}}}
\end{lstlisting}

The compiler is written in the Rust programming language. The function
\texttt{rewrite} (c.f. Listing~\ref{lst:2}) takes as input:
\textcircled{1} the \ac{AST} (\texttt{stmt}), \textcircled{2} a
dynamically growing vector (\texttt{narena}) to allocate the graph
nodes, and \textcircled{3} a vector (\texttt{othreads}) to hold the
initial and ending indices of children threads (those that should run in
synchronous parallel) if any. The function returns the index of the
initial and ending graph nodes for the whole program.

\begin{figure}[tb]
  \centering
  \includegraphics[scale=0.78]{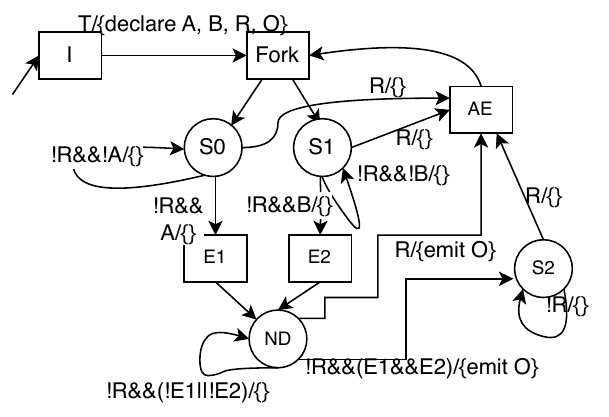}
  \caption{The \ac{FSM} generated for the \texttt{ABRO} example. The
    unneeded dummy nodes are removed.}
  \label{fig:5}
\end{figure}

The algorithm is recursive, matching each kernel statement and producing
a sub-graph following the rewrite rules in Figure~\ref{fig:4}. Consider
the algorithm applied to the \texttt{ABRO} example in
Figure~\ref{fig:2}. Sequential statements
(lines~\ref{re:ln:1}---\ref{re:ln:11}, Figure~\ref{fig:2}) are encountered
first in the \ac{AST}. The algorithm builds the \ac{FSM} bottom up. The
Sequential rule (c.f. Listing~\ref{lst:2}, lines~\ref{seqs}---\ref{seqe})
constructs a graph for the following statement before the preceding
ones. Hence, the graph for the \texttt{loop}
(line~\ref{re:ln:3}---\ref{re:ln:11}, Figure~\ref{fig:2}) will be
constructed first. The \texttt{rewrite} function (Listing~\ref{lst:2},
lines~\ref{loops}---\ref{loope}) first builds the body of the
\texttt{loop}. Same for every other construct, except for
\texttt{pause}, \texttt{nothing}, and signal declarations
(Listing~\ref{lst:2}, lines~\ref{ss}---\ref{se}). Hence, the very first
sub-graph generation occurs for the \texttt{S2:pause} statement on
line~\ref{re:ln:9} in Figure~\ref{fig:2}. Followed by the construction
of the encompassing \texttt{loop} and then the graph for \texttt{emit O}
on line~\ref{re:ln:8} in Figure~\ref{fig:2} and so on and so forth ---
bottom up. The \ac{FSM} generated by applying the \texttt{rewrite}
function in Listing~\ref{lst:2} for the \texttt{ABRO} running example
(c.f. Figure~\ref{fig:2}) is shown in Figure~\ref{fig:5}.

The \ac{FSM} has 4 states, three (\texttt{S0}, \texttt{S1}, and
\texttt{S2}) corresponding to the states in the \texttt{ABRO} example.
The state \texttt{ND} is produced internally by the compiler by applying
the rewrite rule in Figure~\ref{fig:4f}. Unneeded dummy nodes (those
connected to each other) generated by the rewrite rules are removed by a
simple \acf{DFS} of the \ac{FSM}. Only essential dummy nodes
representing fork (\texttt{Fork}), end of parallel blocks (\texttt{E1}
and \texttt{E2}), the exit node due to \texttt{abort} (\texttt{AE}) and
program initialisation node \texttt{I} remain in the generated \ac{FSM}
in addition to the states.

Consider traversing the graph, from the initial node \texttt{I}. First
the signals are declared. Next, the two synchronous parallel blocks are
forked ending up in the \texttt{S0} and \texttt{S1} states together.
This indicates the end of the first clock tick. The \ac{FSM} will remain
in these states until either \texttt{R} happens, in which case the
transition is made to \texttt{AE} and then back into \texttt{Fork},
instantaneously, resulting in restart of these synchronous parallel
statements. If signals \texttt{A} and \texttt{B} are present (and signal
\texttt{R} is absent), then the transition is made to state \texttt{ND}.
From state \texttt{ND}, signal \texttt{O} is emitted, and depending upon
reception of signal \texttt{R} the, \ac{FSM} either restarts the
synchronous parallel statements or halts in state \texttt{S2}. From
\texttt{S2}, the program can only fork the two synchronous parallel
statements upon reception of signal \texttt{R}.

Back-end type-state C++ code can be generated by performing a \ac{DFS}
traversal of the generated \ac{FSM}. The rewrite function in
Listing~\ref{lst:2} and the back-end code generation are both linear in
the number of nodes in the \ac{AST} and the number of nodes and edges in
the \ac{FSM}, respectively. Since \ac{DFS} has a complexity of
$O(|V| + |E|)$, where $V$ and $E$ are the vertices (dummy and state) and
edges in the \ac{FSM}, respectively. Hence, the proposed compilation
time is linear.

\subsection{Back-end code generation}
\label{sec:back-codeg}

Snippets of C++ code generated for the \ac{FSM} in Figure~\ref{fig:5}
are shown in Listing~\ref{lst:3}. The generated C++ adheres to the
tenets of type-state programming. Encoding states as types allows for
reduction in memory footprint in addition to modular formal
reasoning~\cite{garcia2014foundations,field2003typestate}. A number of
optimisations, listed below, are also performed when generating back-end
code.

\begin{enumerate}
\item Lifting signals to the global scope
  (lines~\ref{cpp:1}---\ref{cpp:2}): Signals are declared global and
  \texttt{static}. This allows for using the data section of the
  generated assembler reducing executable code footprint.
\item All states are encoded as empty \texttt{struct}
  (lines~\ref{cpp:3}---\ref{cpp:4}). This should reduce the code size, and
  increase execution speed, since explicit memory is not needed to store
  state of the \ac{FSM} rather it is encoded in the type system.
\item Each synchronous parallel block (thread/statement) is encoded as
  an empty \texttt{struct}, templated over the state following
  type-state programming principles (lines~\ref{cpp:5}---\ref{cpp:6}).
  This again reduces memory footprint and should increase the execution
  speed of the program.
\item Using \texttt{variant} as a type safe union
  (lines~\ref{cpp:7}---\ref{cpp:8}): the \texttt{variant} type is a C++
  construct that allows for aggressive inlining and building jump tables
  in the generated assembler.
\item Using \texttt{inline constexpr} (lines~\ref{cpp:9}---\ref{cpp:10}):
  The C++ compiler not only compiles, but even executes
  \texttt{constexpr} code at \textit{compile} time. This reduces the
  runtime overhead. However, can increase the memory footprint.
\end{enumerate}

\begin{lstlisting}[language=C++, numbers=left, escapechar=|,
  basicstyle=\scriptsize,caption={C++ code snippet generated from the
    \ac{FSM} in Figure~\ref{fig:5}}, label={lst:3}]
// Sig decls
typedef struct signal_A { bool status|\label{cpp:1}| ;} signal_A;
static signal_A A_curr, A_prev; |\label{cpp:2}|....
// Decl states
struct State {}; // generic state |\label{cpp:3}|
struct S2 : State {}; struct S0 : State {};
struct S1 : State {}; |\label{cpp:4}|
// Declare the total number of threads (main)
// and 2 synchronous blocks in ABRO
template <class State> struct Thread0 {}; |\label{cpp:5}|
template <class State> struct Thread1 {};
template <class State> struct Thread2 {}; |\label{cpp:6}|
// Variant decl -- states each thread can be in  |\label{cpp:7}|
using Thread0State =
    std::variant<Thread0<S2>, Thread0<I>, Thread0<ND>>;...
using Thread2State =
    std::variant<Thread2<S1>, Thread2<I>, Thread2<E>; |\label{cpp:8}|
//FSM -- each state transition is encoded as own function
//Thread1 in state S0 function -- same as FSM
inline constexpr void Thread1<S0>::tick(signal_A &_0) { |\label{cpp:9}|
  if ((not R_prev.status) and (not A_prev.status))
    st1 = Thread1<S0>{};
  else if ((not R_prev.status) and (A_prev.status))
    st1 = Thread1<E>{};
  else (R_prev.status)
    st1 = Thread1<E>{};
}.....
static inline __attribute__((always_inline)) constexpr void
visit0(Thread0State &&ts, signal_R &_0, signal_O &_1) {
  std::visit([&_0, &_1](auto &&t) { return t.tick(_0, _1);},
             ts);
}|\label{cpp:10}|...
int main(void) {
  init0();
  while (1) {
    // visit thread 0 -- other threads visited by thread0
    visit0(std::move(st0), R_curr, O_curr);...}
}
\end{lstlisting}

\section{Experimental validation}
\label{sec:experimental-results}

Our over-arching goal is to produce small executables that execute fast,
from imperative synchronous programs, and do so with short compilation
times. In order to validate this goal, we compare the proposed compiler
with freely available, but closed source, official Esterel v6
compiler~\cite{esterelCompiler} and the open source
\ac{CEC}~\cite{edwards2007code}. We call the proposed compiler RUSTC in
the rest of the section. The compiler and benchmarks are available
from~\cite{RUST}.

\subsection{Experimental setup}
\label{sec:experimenal-setup}

\begin{table}[tb]
  \centering
  \caption{Benchmark description}
  \label{tab:1}
  \begin{footnotesize}
    \begin{tabular}{cccc}
      \toprule
      Benchmarks& Code lines & \# threads & \# states\\
      \midrule
      ABRO~\cite{CECBENCH} & 26 & 5 & 3\\
      \midrule
      ABCD~\cite{CECBENCH} & 197 & 6 & 31\\
      \midrule
      MULTI7~\cite{CECBENCH} & 151 & 9 & 33\\
      \midrule
      GRAY\_COUNTER~\cite{CECBENCH} & 191 & 15 & 34\\
      \midrule
      WASH\_MACHINE~\cite{park2015compiling} & 40 & 4 & 23\\
      \midrule
      LEGO~\cite{CECBENCH} & 59 & 4 & 14\\
      \midrule
      MICROWAVE~\cite{CECBENCH} & 59 & 7 & 9\\
      \midrule
      PACEMAKER~\cite{park2014cardiac} & 113 & 11 & 16\\
      \midrule
      ROBOT\_CONTROL~\cite{park2015compiling} & 39 & 7 & 7\\
      \midrule
      FORKLIFT~\cite{CECBENCH} & 439 & 7 & 37\\
      \bottomrule
    \end{tabular}
  \end{footnotesize}
\end{table}

We collect a number of published benchmarks from different sources.
These benchmarks along with their features are listed in
Table~\ref{tab:1}. Standard comparison benchmarks are obtained from the
\ac{CEC}~\cite{CECBENCH} website. The pacemaker, wash\_machine, and
robot\_control benchmarks are obtained from the work of Park et
al.~\cite{park2015compiling,park2014cardiac}. The lines of code, the
number of synchronous parallel blocks (threads) and the number of states
in the program give an indication of the complexity of the program.
Usually the compile time, the generated code size, and execution time
are all affected by the number of states and threads in the program. In
particular, a traditional synchronous programming compiler, generating
automata, at the back-end, will take exponentially long time with
increasing number of threads and
states~\cite{edwards2007code,park2015compiling}, due to the exponential
complexity of the classical synchronous composition.

All our benchmarks are executed on an Intel i7-7600U CPU running at
2.8GHz with 15GB of RAM. The Esterel v6 compiler is a binary download
from~\cite{esterelCompiler}. The CEC compiler written in C++ itself was
compiled from source. Both Esterel v6 and the CEC compiler produce C
code with a single function for the whole program. The C code generated
from the CEC and Esterel v6 compiler is compiled using GCC 15.0 using
-Ofast and -march=native flags. The C++ code generated from RUSTC was
compiled using G++ 15.0 with -Ofast, -march=native, and -std=c++26.
Hence, we are performing a fair comparison between the optimised
executables generated by the three compilers.

\subsection{Experimental results}
\label{sec:experimental-results-1}

\begin{figure}[tbh]
  \centering
  \includegraphics[scale=0.58]{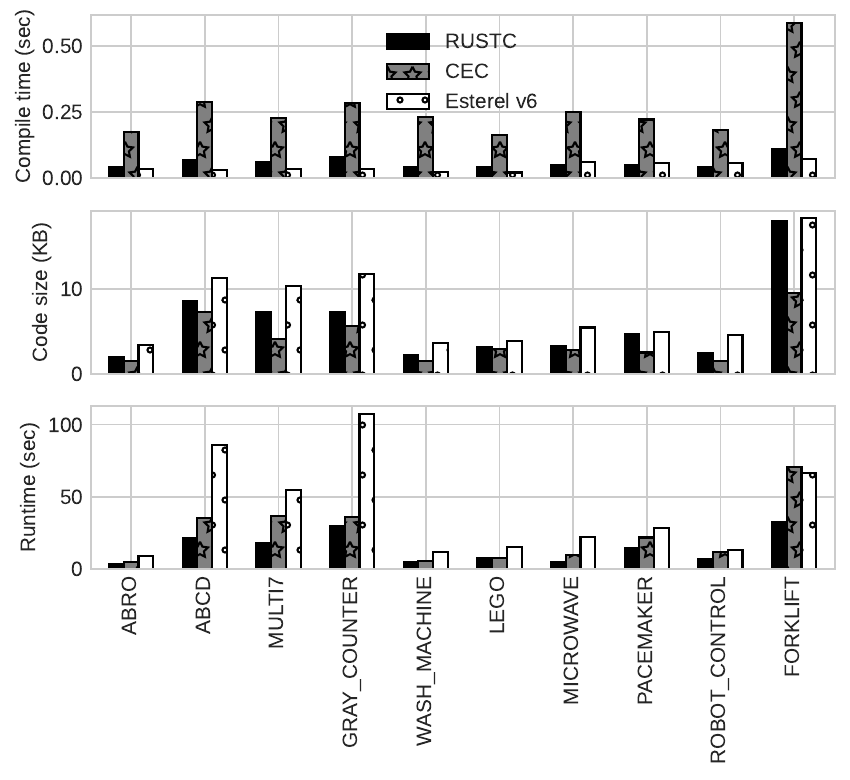}
  \caption{Experimental results --- code size, compile time, and runtime
    comparison. Runtime values are for 1 Billion ticks.}
  \label{fig:res}
\end{figure}

The runtime results for \textcircled{1} compile time, \textcircled{2}
size of the generated code, and \textcircled{3} runtime of the
executables are shown in Figure~\ref{fig:res}. The compile times are
measured using the \texttt{time} command on bash shell. The generated
code size is obtained from \texttt{size} command and includes the size
of the text and data section. The runtimes are obtained by feeding
random inputs to the executable and running the generated program for 1
Billion ticks. The wall clock timer (\texttt{time\_t}) was used to
obtain the runtimes.

As we can see from Figure~\ref{fig:res}, the compile times of the
proposed RUSTC compiler are comparable to the Esterel v6 compiler and
much faster than the \ac{CEC} compiler. On average, the RUSTC
compilation speed is $4.5\times$ faster than \ac{CEC} and $0.69\times$ slower that
the Esterel v6 compiler. The compile time of RUSTC does not grow
exponentially. Recall from Section~\ref{sec:algor-gener-graph}, that our
compilation technique is a \ac{DFS} of the \ac{AST} and the generated
intermediate graph representation of the synchronous program. This
clearly shows that the proposed RUSTC compiler does not suffer from the
state-space-explosion problem like other automata based
compilers~\cite{park2015compiling,potop2007compiling}. The proposed
RUSTC compiler generated binary size, on average, is 29.2\% smaller than
the Esterel v6 compiler and 28.6\% greater than the binary generated by
the \ac{CEC} compiler. This shows that, on the one hand the type-state
programming principles have been effective in reducing the code size. On
the other hand, aggressive in-lining increases the binary size.

The RUSTC compiler outshines both the \ac{CEC} compiler and the Esterel
v6 compiler in terms of execution times of the generated binaries. RUSTC
compiler is consistently better than other compilers in terms of
execution times. On average, the binaries produced by the RUSTC compiler
are 31.6\% faster than \ac{CEC} and 59.9\% faster than those produced by
the Esterel v6 compiler. This increase in execution speed can be
attributed to: \textcircled{1} encoding of program states as types
rather than variables in memory, which reduces the memory accesses and
increases execution speed. \textcircled{2} reduced branching in the
generated code. The \ac{CEC} and Esterel v6 compiler generate switch
case statements for different branches in the program. The proposed
RUSTC compiler on the other hand, uses compile time type inference (with
states encoded as types) along with aggressive inlining to reduce the
number of branches, which leads to reduced branch misprediction and
hence, reduced execution speeds.

Overall, the proposed compilation technique produces very good code
without sacrificing compile times. In fact, the produced executables are
the fastest.


\section{Related Work}
\label{sec:related-work}

Substantial research effort has been dedicated to compilation of
imperative synchronous programming
languages~\cite{potop2007compiling,edwards2007code,edwards1999compiling,yip2023synchronous,park2015compiling}.
The open source \ac{CEC}~\cite{edwards2007code} compiler has been shown
to produce the best code in terms of binary size and execution
efficiency, unto date. However, it is well known that formal
verification becomes harder due to the \ac{CFG} and circuit based
translation of imperative synchronous programs. Recent imperative
synchronous languages such as ForeC~\cite{yip2023synchronous} also take
the approach taken by \ac{CEC} for compiling imperative synchronous
programs to multi-core architectures.

Park et al.'~\cite{park2015compiling} work is the latest in translating
imperative synchronous and \ac{GALS} programs to automata for formal
property based verification. They show that automata based compilation
makes verification of \ac{LTL} properties easier. However, their
compiler takes too long to compile programs with many parallel threads
(blocks) and states, due state space explosion. Hence, their compilation
approach is restricted to only small programs. In the same vein,
automata based compilation for Esterel~\cite{berry1999esterel} was
abandoned due to state space explosion and circuit based compilation has
been adopted by the official Esterel compiler~\cite{berry1999esterel}.

The proposed compilation technique overcomes all these issues. The
proposed compiler produces automata based code at the back-end that is
amenable to modular formal verification~\cite{field2003typestate} of
\ac{LTL} properties, thanks to type-state programming. The compilation
time is linear as it considers every parallel thread independently. The
compilation time and generated executable size is comparable to the
\ac{CEC}~\cite{edwards2007code} and the official Esterel
v6~\cite{berry1999esterel} compilers. Moreover, runtime of the
executable is substantially faster.

\section{Conclusion and Future Work}
\label{sec:concl-future-work}

Synchronous programming languages are prominent in design of safety
critical embedded software. Compiling synchronous imperative programs
efficiently to automata is challenging. This paper proposes a novel
technique for compiling Esterel style kernel synchronous programs into
type-state programs, which enables \textit{modular} compilation and
formal verification. In order to achieve this goal: \textcircled{1}
rewrite rules to construct graphs are introduced, \textcircled{2} a
linear time compilation algorithm applies these rules and reduces the
generated graph into a \acf{FSM}. \textcircled{3} Finally, the \ac{FSM}
is translated in the back-end to a C++ template meta-program encoding
states within the type system.

Experimental results show that the compilation times are comparable to
current state-of-the-art compilers for Esterel and do not suffer from
the state space explosion problem. The generated binary sizes are
comparable to the current state-of-the-art compilers, while being
substantially faster (31--60\%).


In the future we plan to produce automata code for Promela amenable to
the verification via the SPIN model-checker. 
The proposed linear time compilation technique would suit compiling
large and complex imperative synchronous programs for verification.

\bibliographystyle{ACM-Reference-Format}
\bibliography{sample-base}


\end{document}